# Investigating the "*Cocoon Effect*" in Niobium-Copper Alloy: Metallic Nano-Precipitate Distribution and Niobium Migration

*R. L. de Almeida[1,2], J. Albino Aguiar[1] and C. A. C. Passos[2]

[1]Physics Department, Federal University of Pernambuco, 50670-901, Recife, PE, Brazil
[2]Physics Department, Federal University of Espírito Santo, 29075-910, Vitória, ES, Brazil

*Correspondence e-mail: ralmeida@ifi.unicamp.br & almeida.almeida@gmail.com

## ABSTRACT

We report the observation of the metallic niobium migration within the molten Cu-Nb alloy mass on the synthesis of nano-granular Cuxwt%Nb samples. We named "*Cocoon Effect*". To address the complex interplay of this molten phase separation and microstructural evolution, we prepared a series of granular samples by rapidly cooling a molten mixture of Cuxwt%Nb, where the niobium concentration varied (x=3,5,15,20). Our main goal in this work was not only to establish a systematic, innovative and robust method to obtaining good quality samples, but also provide a clear recipe for obtaining similar systems to the investigations of their interesting physical properties. Beyond the understanding of the "*Cocoon Effect*" in Niobium-Copper alloys, we include a wide complementary elsewhere investigation into the very interesting and rich superconducting properties exhibited by the Niobium-Copper alloy. By employing a robust synthesis method, we successfully obtained samples characterized by well-defined spherical nano-precipitates of niobium, featuring regular sizes and grain spacing. Our study contributes not only to our understanding of the Niobium-Copper molten phase separation, microstructure and the "*Cocoon Effect*" in these metallic alloys, but also sheds light on the intricate and important implications for the development and optimization of good quality granular metallic alloys for various applications. From our work, we obtained very impressive microstructural results, such as: $d_m = 1.2\,\mu m$, $D_m < 2.2\,\mu m$ and $\rho = 1.785\,grain/\mu m^2$, where $d_m$ is the distance between Niobium grains, $D_m$ is the mean diameter of Niobium grains and $\rho$ is the Niobium grain mean density in the Copper matrix.

Highlights:
→ Observation and identifiation of the "*Cocoon Effect*".
→ Study of the separation and microstructural evolution of the Niobium-Copper alloys.
→ Determination of the Niobium-Copper alloys microestructural properties.
→ Complementary investigation of the superconducting properties the Niobium-Copper alloy.

# I. INTRODUCTION

The objective of this work was to develop reliable methods for controlled fabrication of materials and nanostructures with varying degrees of complexity for diverse applications [1-6]. Furthermore, there is a continuous pursuit to comprehend the physical properties of these systems, their relationship with structural characteristics, and their reproducibility through simulations [6-9]. Recent advancements in this field have been driven by significant investments made by developed nations in science and technology, with the study of granular materials showing great potential for technological applications [9-12]. Our investigation is a meticulous examination of niobium migration within the molten alloy mass and the our study places particular emphasis on the peripheral region, where the migration of niobium atoms plays a crucial role in the formation of the "*Cocoon Effect*". By tracking and analyzing the migration patterns of niobium, we aim to unravel the fundamental processes and mechanisms that contribute to the observed phenomenon. This investigation not only enhances our understanding of the behavior of niobium and its interactions within the Nb-Cu alloy, but clariry unexplained issues in those systems subject. Furthermore, we aim to explore the implications of the "*Cocoon Effect*" on the microstructural evolution and mechanical properties of the Nb-Cu alloy.

By investigating the distribution and arrangement of precipitates, we can gain a deeper understanding of their influence on the material's strength, toughness, and other mechanical characteristics. This knowledge holds significant potential for optimizing the alloy's properties and tailoring it for specific and applications in industries such metallurgy, material, aerospace, automotive, and structural engineering. Early studies on copper-rich niobium alloys [13] revealed intriguing possibilities for fabricating mechanically reinforced Cu-Nb normal conductors and Cu-$Nb_3Sn$ superconducting wires. The size of niobium precipitates was found to be highly dependent on the cooling rate [14], as the sample temperature decreased from T=1,800°C (liquid) to the peritectic point at T=1,090°C.

In this work, a *Cuxwt%Nb* mixture was melted under T=1,800°C inside a conical radio-frequency coil and dropped inside a water-cooled Cu crucible. The cooling rate was estimated to be $\partial T / \partial t \approx 2,800 °C/s$, producing a regular distribution on Niobium precipitates nearly of spherical shapes, with diameters typically around or below 1,000*nm*. The pellets were initially suspended by a thin tungsten wire and dropped down when a homogeneous liquid phase was attained. Plate-like samples having typically an area of 10$mm^2$ and 0.1*mm* in thickness were then obtained. To study the Niobium-Copper system, we prepared rectangular pieces (5.0×1.5×0.1 $mm^3$) from a considerable number of good samples from materials with high purity, Cu (BCC) ≈0.99999 and Nb (FCC) ≈ 0.99999, with different concentrations of Nb, that is, (x=3,5,15,20) in weight (mass) percent. The preparation was done in three phases. The (FCC)-Cu matrix (BCC)-Nb precipitates with the use of these techniques have sizes and regular spacing and very suitable spheroidal geometry. Structural characterization was made by Optical Microscopy and Scanning Electron Microscopy (SEM-JEOL$^{Ltd}$). The unshown results of the experimental dc measurements $M^{dc}[H,T,\theta]$ - (MPMS®@ $H_{ext}^{max}$ = 50 kOe), ac $\chi^{ac}[T,H,h_0,\upsilon,\theta]$, as well the magneto-transport measures $\rho^{dc}[H,T,\theta]$, $R^{dc}[V,IH,T,\theta]$, $I[V,H,T,\theta]$ and $J_c[H,T,I,\theta]$ - (PPMS® @ $H_{ext}^{max}$=90 kOe) will be discussed soon elsewhere.

## II. PROCEDURE AND EXPERIMENTAL DETAILS

## II.I. THE NIOBIUM-COPPER MICROSTRUCTURE EVOLUTION

We prepared a considerable wide number of *CuXwt%Nb* samples with x=3,5,10,15 and 20, from Copper and Niobium materials with high purity, e.g., Face-Centered Cubic Lattice (Cu)≈ 99.999% and Body Centered Cubic (Nb)≈ 9.99%.

The preparation of these samples was particularly difficult because high difference between the melting points of Copper 1,085 oC and Niobium, above 2,477 oC and, from the melting temperature during the cooling, there is a strong tendency of the formation of other phases between these two materials [14].

The first one phase diagram for Cu-Nb alloys was constructed in [13] and the second one, more complete, which includes two phase transformations independent of temperature, that is, a monothetic in approximately $T_{monothetic}^{(Cuxwt\%)} \approx 1.545$ oC and a perithetic reaction in $T_{perithetic}^{(Cuxwt\%)} = 1,083$ oC was showed in [15]. At this phase diagram [15], there are three phases strongly dependent on temperature. The first phase $\alpha + \beta$, which lies in an isotherm immediately below $T_{perithetic}^{(Cuxwt\%)} = 1,083$ oC (perithetic region) consists of both constituents of the alloy in its solid phase. The phase $L + \beta$ which is situated above T=1,083 oC and a non-isothermal line that goes up to the melting temperature of Nb at T≈2,477 oC, depending on of the concentration x in the Cuxwt%Nb alloy. It is consisting of copper in the liquid phase and niobium in the solid phase. This phase has a species of plateau (monothetic region) at $T_{monothetic}^{(Cuxwt\%)} = 1,545$ oC. The third and final phase, the $L - phase$ which is situated above the line delimiting the phase $L + \beta$, and the region of the diagram where all two constituents of the material are in their liquid phase (depending on the concentration of Nb in Cu). The microstructure of the samples based on Cuxwt%Nb it's strongly dependent on the cooling rate $\partial T / \partial t$ of system [14,-16-18] and, we verify through that, how faster the molten CuXwt%Nb system is brought to thermal equilibrium with the room temperature, better is the distribution and organized size of Nb particles in the Cu matrix. The solubility of Nb in Cu $S_{Nb \to Cu}$ is around $1 wt \%$ at a temperature below $T_{1wt\%}^{(S_{Nb \to Cu})} = 1,100$ oC, and this solubility decreases very rapidly with the reduction of the temperature and, at room temperature, this same solubility $S_{Nb \to Cu}$ drops to a value less than $0.1 wt \%$, $T_{0.1wt\%}^{(S_{Nb \to Cu})} = 300$ K [16-18]. In materials cooled in rates close to those achieved in this work, this solubility $S_{Nb \to Cu}$ does not exceed $0.15 wt \%$. On the other hand, the border of the Nb is little known, however, there are experimental data [18] what indicate that the solubility of Cu in Nb $S_{Cu \to Nb}$ at temperatures close to $T_{0.07wt\%}^{(S_{Cu \to Nb})} = 1,100$ oC is around $0.07 wt \%$. In this work, to prepare the samples, the materials were weighed, and the concentrations x in samples with stoichiometry Cuxwt%Nb, was varied between x = 3,5,10,15 and 20.

The subsequent step was to melt the constituent materials (Cu) and (Nb) that were taken into the of an arc furnace chamber $I_{T=300K}^{max} = 500$ A. After this step, we used a *RF* furnace in a "molten mass levitation configuration" to redistribute the Nb grains at the Cu matrix. A new and interesting effect we observe during the melting of the samples in the *RF* oven in an almost levitation configuration, it was the tendency of niobium to migrate to the peripheral regions of the samples. And there was a kind of involvement of all the melted mass by a thin Nb film. We labeled this effect as a "*Cocoon effect*" because it was in fact a kind of Niobium "Cocoon" that formed in around the melted Copper melted mass during the melting of the samples in the RF furnace. We identified the sample regions where there was the highest Nb concentration, due to the diffusive effect during the of *RF* melt process. For that, in a part of the sample labeled as Cuxwt%Nb-11-02. We identify three regions and separated them and we put the electron beam spot of the scanning electron microscope separately in each of these regions, to obtain more information (chemical and compositional) on the microstructure of our samples from each region.

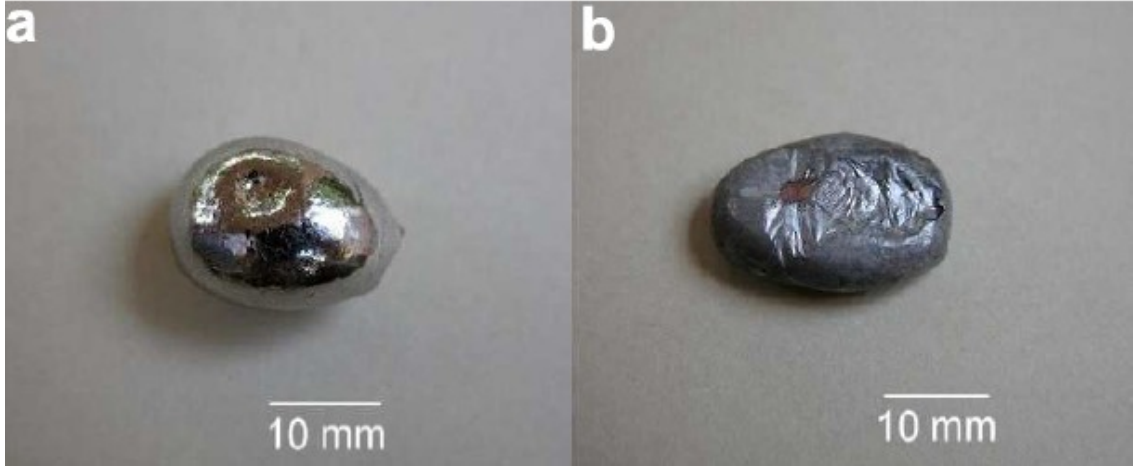

Figure 1: Samples Cuxwt%Nb-11-02 (a) and the sample Cuxwt%Nb-10-01 (b) immediately after they are melted in the arc furnace. If we look closely at the sample Cuxwt%Nb-11-02 (a) we perceive a typical brightness that is not characteristic of Cu (even this sample has x=80 % by weight of Cu). Microscopy images scanning revealed the migration of the Nb to the peripherical regions of the sample: "C*ocoon Effect*".

We estimate the cooling rate of the system and its limits, which take into account the measured down time (very short $\tau \approx 0.3\,s - 0.5\,s$) for the sample drop from the RF coil to the inside of the refrigerated Cu crucible based on the assumption that the emission of thermal energy through radiation by the sample and by heat transfer between the sample and the crucible to a temperature less than or equal to T=1000 $^o$C, where both elements are already found in the solid phase. We used the coefficient of Copper thermal conductivity, which is 3.8 J/cm² $^o$C, and, from this measure made several times for different samples niobium content, we determine the rate in two stages: The first exclusively by radiation, taking in the emission of radiation between two bodies:

$$Q_{12} = F_{12} \sigma T_1^4 A_1 \qquad (1)$$

where, $Q_{12}$ is the thermal energy from the body 1 (in if the sample flows into the entire inner chamber of the chamber, which is the body 2), $F_{12}$ is the superficial fraction of any and all the receiving area $A_1$ of energy seen by the area that is emitting (molten sample). It is valid to consider that for radial direction to the heat path (which is the case with our system – "spheroidal" geometry). So, $F_{12}=1$, T is the sample temperature and, the temperature $\sigma$ is the Wien constant (or Stephan-Boltzmann constant). And, to estimate with greater, as much as possible, the rate of Fourier thermal conductivity equation

$$Q_{12} = - KA \left( \frac{\partial T}{\partial l} \right) \qquad (2)$$

In the equation 2, *K* is the thermal diffusivity constant, in this case, for the Cu, which is the dominant quantity material in the sample, and A the effective contact area between the two bodies. Overlapping equations 1 and 2 and, with some small and easy algebraic work, we estimate the molten sample mass cooling rate determined by the final expression:

$$\frac{\partial T}{\partial l} = \frac{\Delta T\,F_1 + \Delta T\,F_2}{\Delta l\,F_1 + \Delta l_{F2}} \approx 2{,}800\,K/s \qquad (3)$$

In the figure 2 we showed optical microscopy photos from the samples after cutting and polishing to carry measurements on the scanning electron microscopy. The sample Cuxwt%Nb-11-02 is shown in letter *a*, and sample Cuxwt%Nb-10-01 is shown in the letter *b* (both in figure 2). We added that this cutting geometry was chosen only for the accomplishment of the structural characterization through the images of *Scanning Electron Microscopy*. For the measures of the superconducting properties (inductive and magneto-transport), the samples were prepared in the form of stripes, with dimensions at least five times smaller than the shown in the figure 2 (see geometric dimensions of stripes, de *Cuxwt%Nb* in the table 1). Those data will not be showed here. Sample:

| Cuxwt%Nb Samples Properties | | | | |
|---|---|---|---|---|
| Samples | Dimensions *lxwxt* | Nb: Xwt% | Mass(g) | Geometry |
| CuXwt%Nb-**10-01** | 0.50x0.15x0.016cm$^3$ | X= 10 | (0.270$\pm$0.0001) | Rectangular Stripe |
| CuXwt%Nb-**10-02** | 0.40x0.20x0.01cm$^3$ | X= 20 | (0.210$\pm$0.0001) | Rectangular Stripe |

Table 1: Table containing information about structural properties of the samples CuXwt%N. We will concentrate our discussions on the two samples mentioned in this table (Cuxwt%Nb-10-01 e Cuxwt%Nb-11-02).

In the figures 3 and 4 we show the scanning electron microscopy images of both samples *Cuxwt%Nb*. As we can see, the geometry definition of Nb grains obtained in the copper matrix is much higher than that we had obtained the sub-micrometric grains of Nb at the previous mentioned references. However, we still had the task of re-distribute better the Niobium nano-grains throughout the volume of the Cu matrix. Thus, the last melting stage of the *Cuxwt%Nb* system sample preparation, consisted of the use of a resistive furnace for a annealing of the sample system, with the objective of reaching only melting temperature of copper, $T=1,083°C$, and by the diffusion effects, since there is a considerable density difference between Cu and Nb of $D_{Nb} - D_{Cu} = 0.56 \, g/cm^3$.

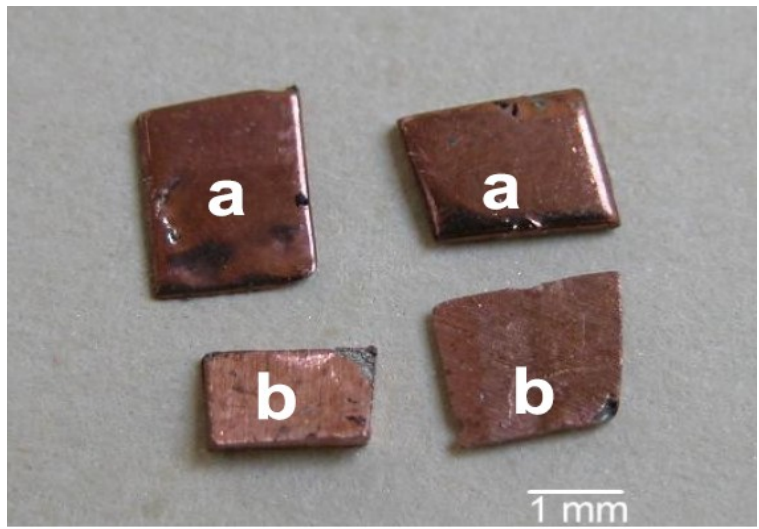

Figure 2: Samples *Cuxwt%Nb*-11-02 (a) and *Cuxwt%Nb*-10-01 (b) to be measured in the scanning electron microscope.

Thus, after a very smooth climb ramp temperature, $\Delta T/\Delta t = 100°C/h$, we raise the system to an equal temperature a 1,100 $^o$C. We let the system stay in this temperature for one to two hours. After that, we reduced temperature on a faster ramp $\Delta T/\Delta t = 250°C/h$. The results obtained with the spacial "rearrangement" distribution of the Nb grains in the Cu matrix can be seen in the figures 5 and 6 for the sample Cuxwt%Nb-11-02, that is much better distribution once compared with the grain distribution on the figures 3 and 4. How can be observed in the figures 5 and 6, the distribution of Nb grains in the copper matrix is quite adequate, as there is a very good regularity in the spacing between the grains of Nb in the Cu matrix, and again, we did not find results similar to these in the previous literature.

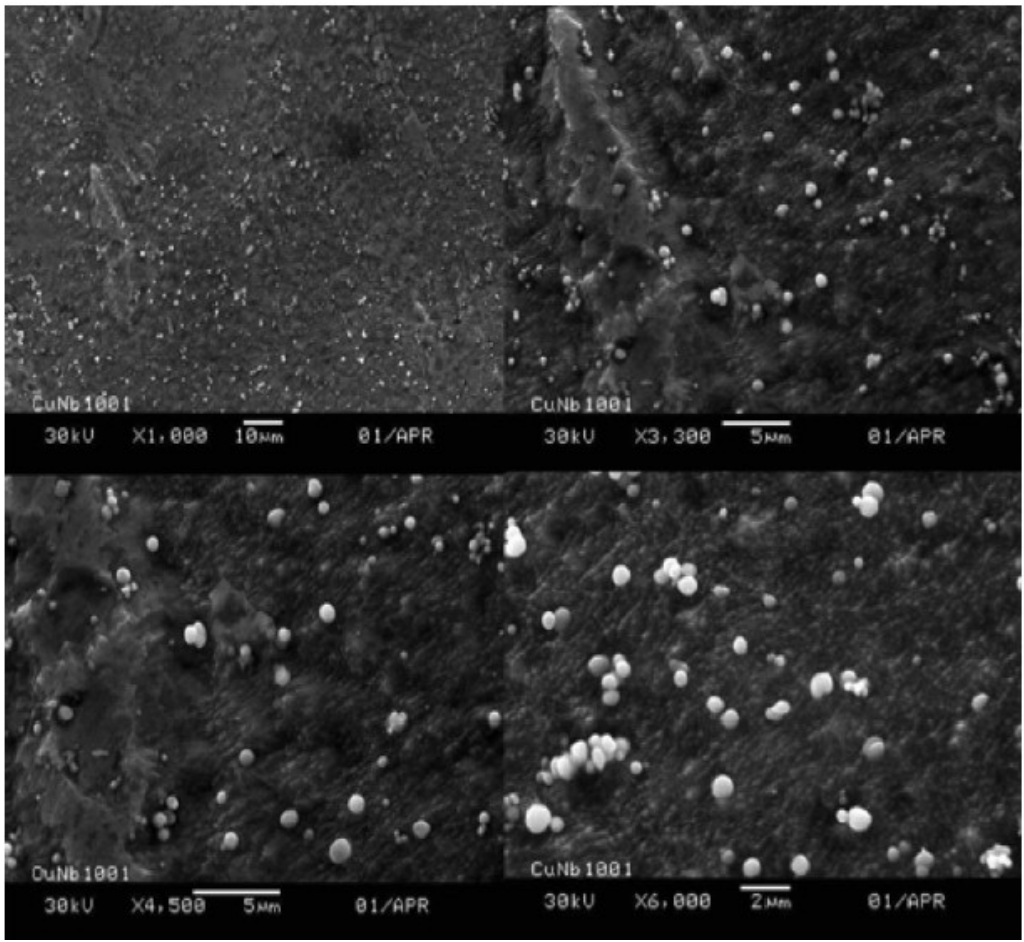

Figure 3: Scanning Electron microscopy images from sample Cuxwt%Nb-10-01). As can be seen, the beam energy used for this measure was $E_{el}=30\,kV$, the mode of obtaining the images was the backscattered electrons The grains of Nb in the Cu matrix have dimensions smaller than $1\,\mu m$.

The geometry and granular spacing remain constant in all alloy surface. The dark spots on some grains of Nb are due to propanone residues ($C_4H_8O$) what is it used to clean up the surface of the samples prior to realization of the images, because. The adopted measurement method was the backscattered electrons method, it's that in this way, it "favors" the contrast image from elements with greater atomic number, and the carbon, the main constituent of the ketone, has an atomic number smaller than the other two elements on the sample, and therefore the dark shadow color on some grains of Nb. From the results of the last annealing fusion, we made a statistical analysis of the sample Cuxwt%Nb- 11-02, using the scanning microscopy measurements, and we obtained the results that are found in the graphic of the 7 and 9.

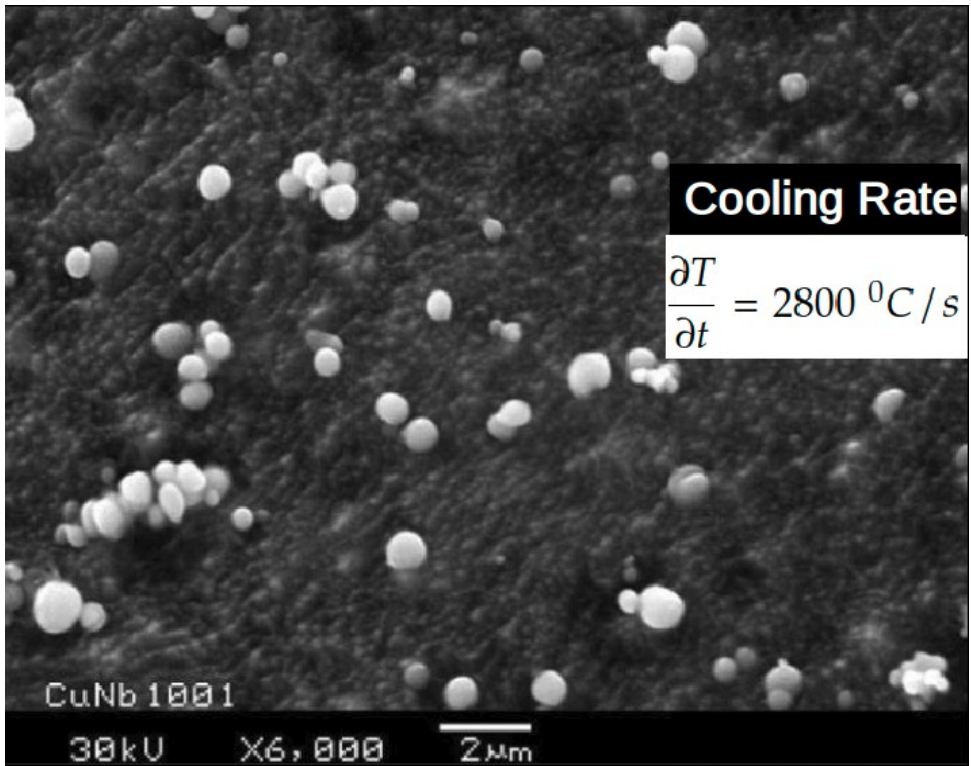

Figure 4: Electron microscopy image of scan showing the microstructure of the sample in more detail Cuxwt%Nb – 10-01. $E_{el}=30\,kV$. The mode of obtaining the images was that of backscattered electrons, and the size of the grains of Nb in the matrix is around $1\,\mu m$.

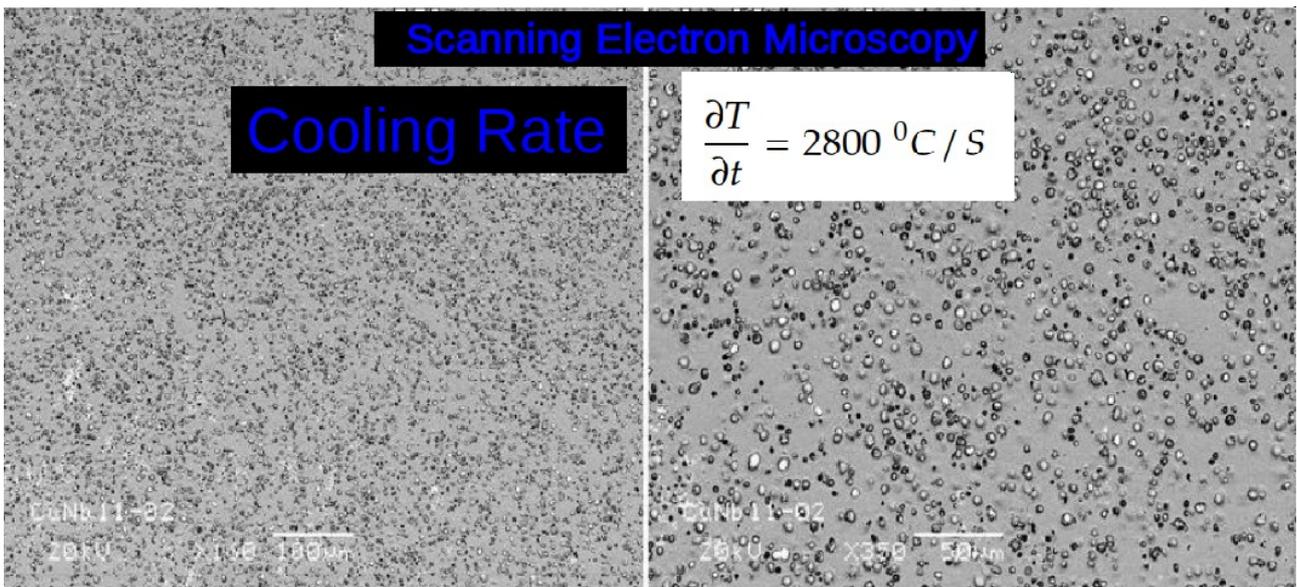

Figure 5: Scaning Electron Microscopy images of showing the sample microstructure Cuxwt%Nb – with a very better Nb grains distribution along of the matrix.

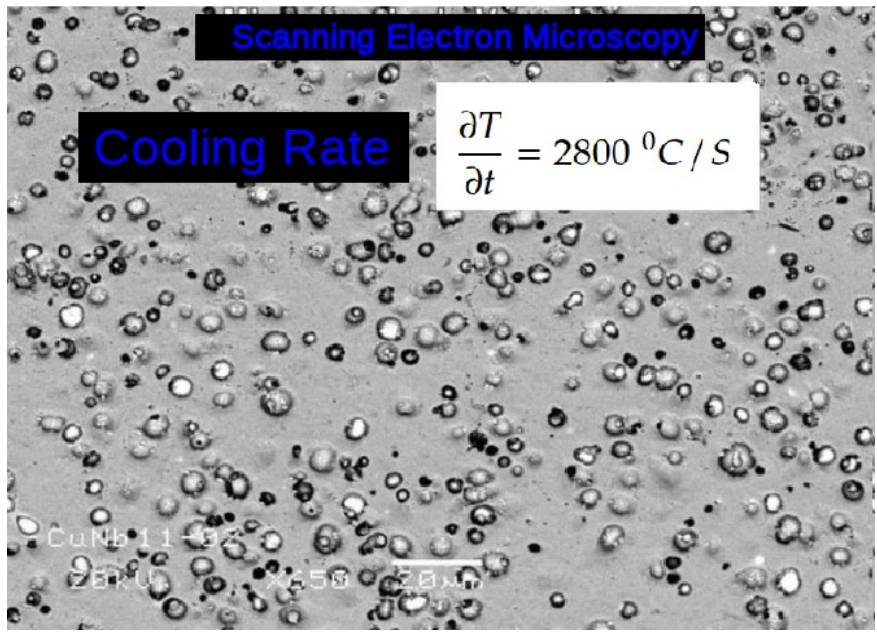

Figure 6: Electron microscopy image of with the microstructure of the sample Cuxwt%Nb-11-02 in details.

After completing the first stage, which consisted of preparation (growth in three merger processes), microstructural characterization by Scanning Electron Microscopy-(SEM-JEOL[Ltd]), e.g., qualitative and quantitative Energy-dispersive X-ray spectroscopy/Energy dispersive X-ray analysis (*EDX* and *EDS*).

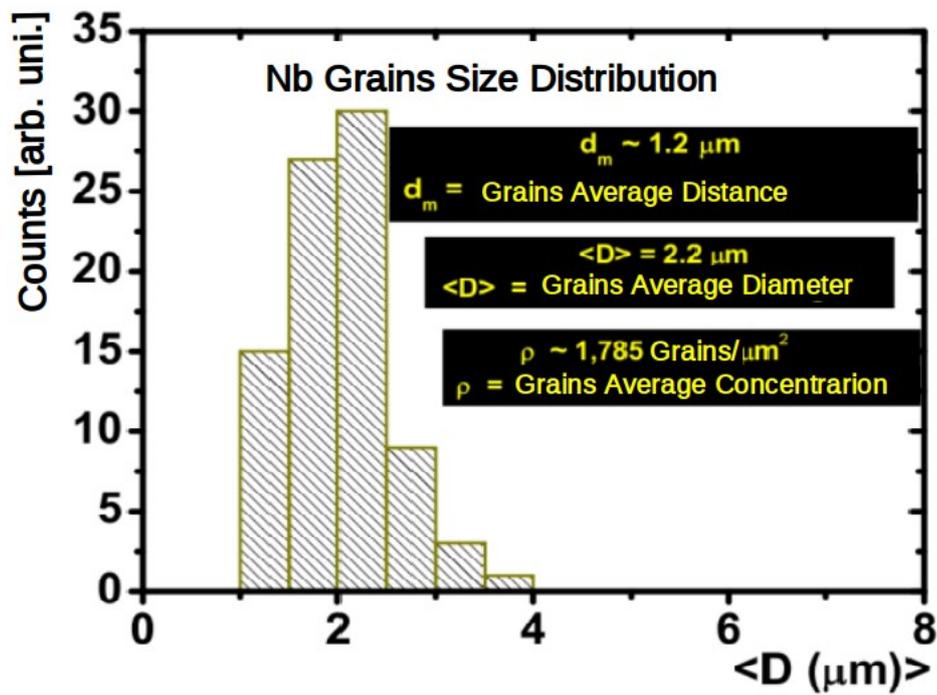

Figure 7: Quantitative Statistical Analysis from the microstructural properties from the sample CuXwt%Nb-11-02 carried out through the data obtained from the images of scanning electron microscopic.

As shown above, we will show the physical characterization properties for the samples Cuxwt%Nb (x=3,5,10,15 and 20), with focus for the concentrations x=10% and 20%. We will show below theresults of $M^{dc}[H,T,\theta]$, $\chi^{ac}[T,H,h_{Exc},\upsilon,\theta]$, $\rho^{dc}[H,T,\theta]$, $R^{dc}[V,IH,T,\theta]$, $I[V,H,T,\theta]$ and $J[H,T,I,\theta]$ ]characterization of the superconducting properties of these granular samples.

## II.II. The Niobium-Copper *Cocoon Effect*

A very interesting effect not yet reported in the scientific literature that we observed during the melting of samples in the RF oven in a levitation configuration, was the tendency of niobium molten liquid mass to migrate diffusively to the peripheral regions of the samples. And there was a kind of involvement of the whole melted sample volume by a Niobium thin film (see Fig 1). We labeled this effect the effect "*Cocoon Effect*" because it was, in fact, a kind of Nb "cocoon" that was formed around the entire melted mass (Niobium+Copper) during samples melting process in the RF furnace coil at the levitation configuration. We identified the regions of the Figure 8.

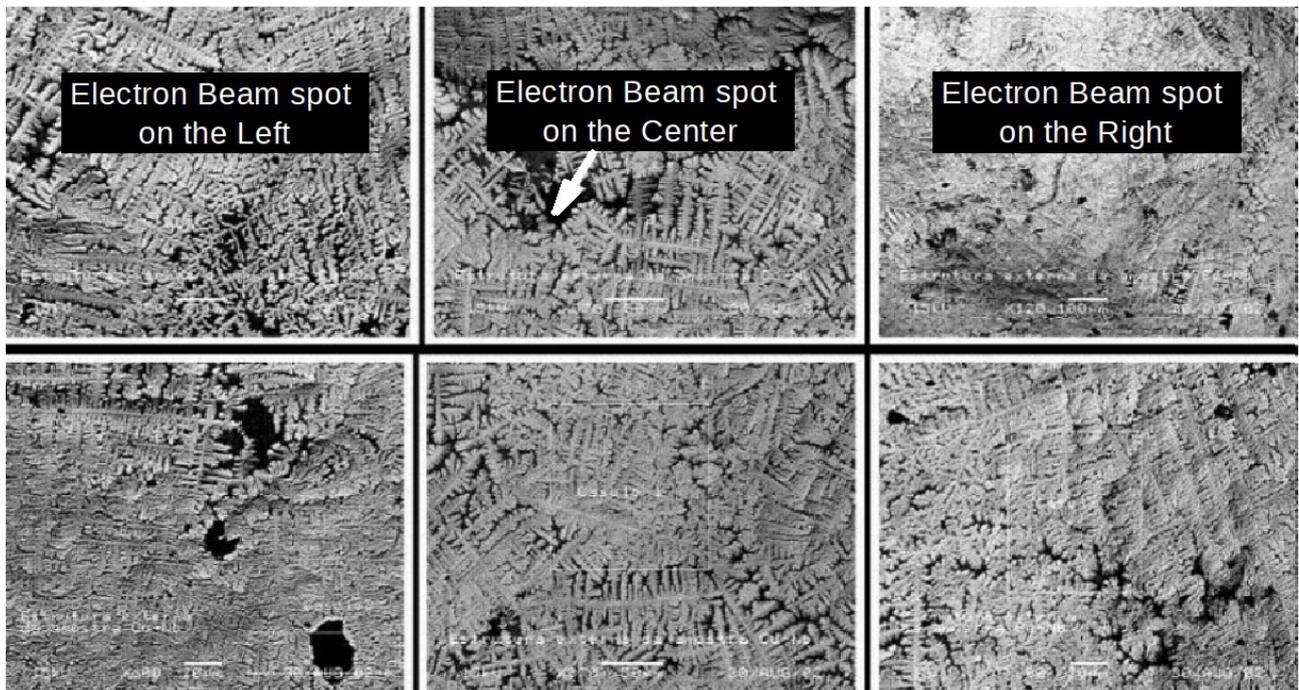

Figure 8: Scanning Electron Microscopy image of with the microstructure of the sample Cuxwt%Nb-11-02 in details of the "*Cocoon Effec*t".

The regions of the sample CuXwt%Nb-11-02 was selected to be analyzed qualitatively with measurements energy dispersive (EDX/EDS) that are shown in figure 8. The samples where there was the highest concentration of Nb, due to the diffusion effect during theRF fusion process. Therefore, in a part of the sample CuXwt%Nb-11-02 (whose properties are mentioned in table 1), we identified three regions, separated them and focus the scanning microscope electron beam spot to get more information (chemical and compositional) from the microstructure of the samples. The graphs with such results are shown in figure 7 and 9. These results have again the migratory effect of Nb to the ends of the samples during the fusion. We estimate the cooling rate of the system, which take into account the falling time (very short ~ 0.03s-0.05s) of the sample, the coil to the interior of the cooled Cu crucible, assuming that the emission of thermal energy through radiation by the sample and through heat transfer between the sample and the crucible to a temperature lower than or equal to 1000 ºC, where both elements elements are already in the solid phase, we use the conductivity coefficient temperature of Cu, which is 3.8 J/cm$^2$ ºC.

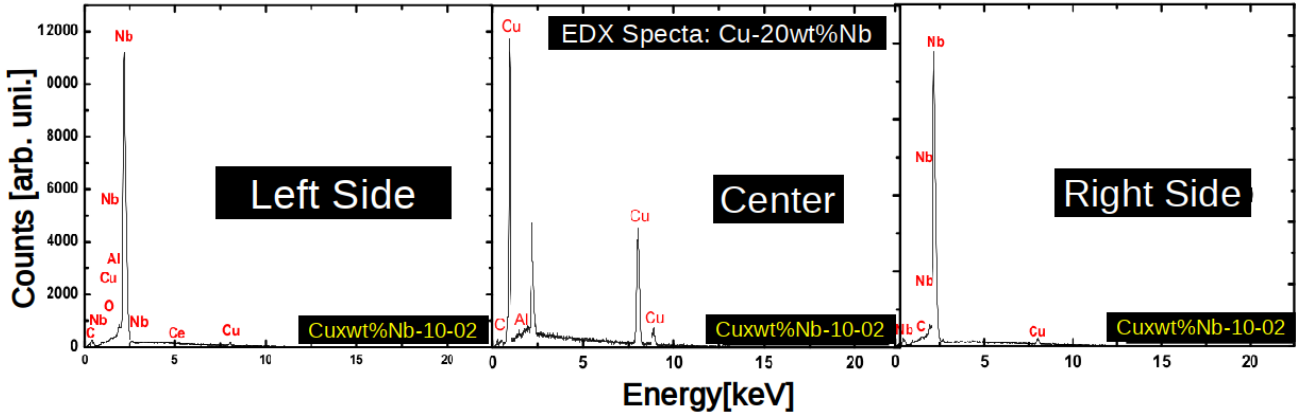

Figure 9: Dispersive energy measurements (EDX) for the sample CuXwt%Nb – 11- 02. Measurement performed with the spot of the electron beam on the: a) right b) center c) left.

Through our analysis from the our results, we obtained the following microstructural data results from the figures 2, 4, 5, 6, 7 and 9:
- $d_m = 1.2\,\mu m$, where $d_m$ is the distance between grains.
- $D_m < 2.2\,\mu m$, where $D_m$ is the mean diameter of Niobium grains.
- $\rho = 1.785\,grain/\mu m^2$, where is the mean grain density of Nb in matrix of Cu.

After finishing the first stage, which consists of preparation (growth in three fusion processes), microstructural characterization (qualitative and quantitative, EDX/EDS data - SEM-JEOL$^{Ltd}$) as shown in this work, from CuXwt%Nb samples (X=3, 5, 10, 15 and 20), we will show elsewhere superconducting properties characterization results of the of these samples granular, with the following measures: $dc$ inductive experiments, $M^{dc}[H,T,\theta]$ on a Magnetic Property Measurement System (MPMS® with Superconducting Quantum Interference Device-SQUID detection @ $H_{ext}^{max}=$ 50 kOe), ac inductive magnetization $\chi^{ac}[T,H,h_{Exc},\upsilon,\theta]$, magneto-transport measures, $\rho^{dc}[H,T,\theta]$, $R^{dc}[V,I\ H,T,\theta]$, $I[V,H,T,\theta]$ and $J[H,T,I,\theta]$ using a micro Hall sensor with 10 measuring transverse voltage bridges regularly distributed throughout the longer sensor length, with dimensions $5.0\times0.2\times0.01\ mm^3$ at the platform Physical Property Measurement System (PPMS® @ $H_{ext}^{max}=90$ kOe). Those results will be published soon.

## III. CONCLUSION

In this study, we have successfully developed a simple, robust, and highly reproducible method for preparing a series of nano granular metallic samples within the Cuxwt%Nb system, with a specific focus on the stoichiometries Cu10wt%Nb and Cu20wt%Nb. Through the utilization of SEM-JEOL$^{Ltd}$, we have made a complete characterization of the mopholoogic/structural properties of these samples, enabling a comprehensive understanding of their microstructural features.

To further enhance the arrangement of niobium (Nb) grains within the copper (Cu) matrix, we implemented an additional step after RF melting, which we refer to as matrix annealing. This step effectively improves a very good rearrangement of Nb grains within the Cu matrix, leading to an optimized granular structure with enhanced superconducting properties. The developed methodology provides a reliable and reproducible route for fabricating granular superconducting samples in the Cuxwt%Nb system, facilitating future investigations into their physical and electrical properties. The SEM characterization allowed us to confirm the successful formation of well-defined Nb grains within the Cu matrix, further validating the effectiveness of the fabrication method. The improved arrangement of Nb grains within the Cu matrix, achieved through the matrix annealing step, holds promise for enhancing the physical properties of the system. In order to improve even more the arrangement of Nb grains in the Cu matrix, we submit the system to a third step, after RF melting, which we define as an improvement of grain rearrangement in the matrix annealing. This step consisted of placing the system in a resistive raise the system close to Cu melting temperature $T_{raise}^{(Cu)} = 1100^oC$ (Cu melting temperature Cu $T_{melting}^{(Cu)} = 1083^oC$) and leave the system remain in this state for 48 hours until by difference of density and performance of viscous (and diffusion) forces, the grains of Nb to reach the ideal dispersion configuration in the Cu matrix. The precipitates of Nb that we obtained with the use of these techniques, have average sizes $\langle D_m \rangle \approx 1.0475 \mu m$, spacing between the grains $\langle d_m \rangle \approx 1.20 \mu m$, $\rho = 1.785 grain/\mu m^2$ (mean grain density of Nb over the Cu matrix, spheroidal shape geometry "well behaved" and very good quality. Finally, our study has successfully developed a reproducible method for preparing granular superconducting samples in the Cuxwt%Nb system. The SEM characterization confirmed the formation of well-defined Nb grains within the Cu matrix, and the matrix annealing step further improved the arrangement of these grains.

## IV. FINAL REMARKS

In a complementary work, we established the complete correlation of the microstructural properties of these good quality *Cuwt%Nb* samples, with their superconducting properties through huge set of complementary measurements, e.g., *dc* inductive experiments, $M^{dc}[H,T,\theta]$, $\chi^{ac}[T,H,h_{Exc},\upsilon,\theta]$, $\rho^{dc}[H,T,\theta]$, $R^{dc}[V,I\ H,T,\theta]$, $I[V,H,T,\theta]$ and $J[H,T,I,\theta]$ and we have observed effects of quantum interference in the critical current curves $I_c[H]$ or $[H,T,I,\theta]$) with the occurrence oscillations due the Josphson Effect coupling betwen the superconducting phase form the Niobium grains.

## V. AKNOWLEDGEMENTS

We would like to thank Professor Oscar Ferreira de Lima and Dr. Adelino Coelho form the "Gleb Wataghin" Physics Institute at the State University of Campinas – Unicamp. We also thank the Brazilian Synchrotron Light National Laboratory (LNLS) - ABTLuS - Ministry of Science, Technology and Innovation – (MCTI) and the National Council for Scientific and Technological Development, CNPq.